\documentclass{aa}
\usepackage{astrobib,epsfig}
\begin{document}
%
%
\def\THCI {\hbox{$^{13}$C{\sc i}}}
\def\C   {\hbox{$^{12}$C}}
\def\CI  {\hbox{$^{12}$C{\sc i}}}
\def\THC {\hbox{$^{13}$C}}
\def\tpo {\mbox{$^3P_1\rightarrow $$^3P_0$}}
\def\tpt {\mbox{$^3P_2\rightarrow $$^3P_1$}}
\def\THCO {\hbox{$^{13}{\rm CO}$}}
\def\HII {\hbox{H~{\sc ii}}}
\def\THCEIO {\hbox{$^{13}{\rm C}^{18}{\rm O}$}}  
\def\THCp {\hbox{$^{13}$C$^{+}$}}
\def\TWCO {\hbox{${\rm ^{12}CO}$}} 
\def\Cp  {\hbox{$^{12}$C$^{+}$}}
\def\CEIO {\hbox{${\rm C}^{18}{\rm O}$}}  
\def\Lbol {$\hbox{L}_{bol}$}
\def\Lsol {$\hbox{L}_\odot$}
\def\kms{\,km~s$^{-1}$}  
\def\TRST {${T_R}^*$}
\def\etal {{et~al.}\ }
\def \al {$\alpha $}
\def\eq {equation}
\def\CII {\hbox{C~{\sc ii}}}
\def\micron{\hbox{$\mu$m}}
\def\OI  {\hbox{O{\sc i}}}
\def\MOLH {\hbox{${\rm H}_2$}}  
\def\perccm {\hbox{${\rm cm}^{-3}$}}   
\def\perscm  {\hbox{${\rm cm}^{-2}$}}
\def\apj{{ApJ} }
\def\aua{{A\&A} }
\def\auas{{A\&AS} }
\def\apjs{{ApJS} }
\title{\THCI\ in High-mass Star-forming Clouds}
\author{ A. R.~Tieftrunk\inst{1} \and K.~Jacobs\inst{1} \and C.
  L.~Martin\inst{2} \and O.~Siebertz\inst{1} \and\\ A. A.~Stark\inst{2} \and
  J.~Stutzki\inst{1} \and C. K.~Walker\inst{3} \and G. A.~Wright\inst{4}}
\offprints{A.R.~Tieftrunk \email{atieftru@ph1.uni-koeln.de}}
\institute{KOSMA, I. Physikalisches Institut der Universit\"at K\"oln,
  Z\"ulpicher Str. 77, 50937 K\"oln \and Smithsonian Astrophysical
  Observatory, 60 Garden Street, Cambridge MA 02138, USA \and Steward
  Observatory, University of Arizona, Tucson AZ 85721, USA \and Bell
  Laboratories, 791 Holmdel-Keyport Rd., Holmdel NJ 07733, USA
  }
\date{Received 10.05.2001 / Accepted 27.06.2001 }
%
%
\abstract{ We report measurements of the \C/\THC\ abundance ratio in the
  three galactic regions G 333.0-0.4, NGC 6334 A and G 351.6-1.3 from
  observations of the \CI\ \tpt\ transition and the hyperfine components of
  the corresponding \THCI\ transition near 809 GHz. These transitions were
  observed simultaneously with the CO 7--6 line emission at 806 GHz with the
  AST/RO telescope located at the South Pole.  From a simultaneous fit to
  the \CI\ \tpt\ transition and the HF components of the corresponding
  \THCI\ transition and an independent estimate of an upper limit to the
  optical depth of the \CI\ emission we determine intrinsic \CI/\THCI\ 
  column density ratios of $23\pm 1$ for G 333.0-0.4, $56\pm 14$ for NGC
  6334 A and $69\pm 12$ for G 351.6-1.3.  As the regions observed are photon
  dominated, we argue that the apparent enhancement in the abundance of
  \THC\ towards G 333.0-0.4 may be due to strong isotope-selective
  photodissociation of \THCO, outweighing the effects of chemical isotopic
  fractionation as suggested by models of PDRs. Towards NGC 6334 A and G
  351.6-1.3 these effects appear to be balanced, similar to the situation
  for the Orion Bar region observed by Keene et al. (1998).
}
\authorrunning{A. R. Tieftrunk et al.}
\titlerunning{\THCI\ in High-mass Star-forming Clouds}
\maketitle
\section{Introduction}

The study of CNO isotope abundance ratios in the interstellar medium is 
crucial to understanding galactic chemical evolution.  In particular, the
ubiquity of C-based molecules in interstellar clouds has made the
\C/\THC\ abundance ratio an important chemical diagnostic. From measurements
of \CEIO\ and \THCEIO\ in 13 interstellar clouds, \citeANP{LP90}\ 
(\citeyearNP{LP90}, \citeyearNP{LP93}) found a galactic gradient in the
local carbon isotope abundance ratio (hereafter: n(C)-ratio) ranging from 25
towards the galactic center to about $60-70$ in the local ISM and out to a
galactic radius of about 10 kpc. They also found an increase in the
n(C)-ratio derived from CO towards clouds exposed to higher UV radiation
fields (namely Orion KL and W33), supporting the
suggestion made by models of photon dominated regions (PDRs) that depletion
in \THCO\ can be ascribed to inefficient self-shielding resulting in
isotope-selective photodissociation \cite{DB88}.

As discussed by \citeN{KSK98}, who report the first detection of the
strongest of the \THCI\ hyperfine (HF) structure components at 809 GHz from
the Orion Bar region, the abundance of atomic carbon is sensitive to the
effects of chemical isotopic fractionation and isotope-selective
photodissociation, allowing the study of the significance of these competing
effects in PDRs and to verify model predictions (e.g.  \citeNP{LPR93},
\citeNP{KSS94}).  The charge exchange fractionation reaction, \THCp\ $+$
\TWCO\ $\rightleftharpoons$ \Cp\ $+$ \THCO\ $+$ 36~K, being exothermic,
preferentially incorporates \THCp\ into \THCO, thus leading to a local
\THCp\ and \THCI\ depletion and a corresponding \THCO\ enhancement.  This
may be balanced by isotope-selective photodissociation, which in contrast
reduces the gas-phase abundance of the less efficiently self-shielded \THCO\ 
isotopomer \cite{DB88}. Models show that the proportion of these effects
strongly depends on the temperature derived for the C$^+$/C{\sc I}/CO
transition zone (due to the low energy barrier of the exchange reaction) and
the comparison between different models gives non-conclusive results
(\citeNP{KSS94}; \citeNP{LPR93}).

From their measurement of the \CI\ \tpt\ and the strongest HF component of
the \THCI\ equivalent transition, \citeN{KSK98} find a N(\C)/N(\THC) column
density ratio (hereafter: N(C)-ratio) of $58\pm 12$ towards a position near
the western end of the Orion bar.  They derive a somewhat higher ratio of
$75\pm 9$ towards the same position from observations of \CEIO\ and \THCEIO.
In comparison, direct observations of the \Cp/\THCp\ abundance ratio
(\citeNP{SJG93}; \citeNP{BB96}), which should not be significantly
influenced by chemical isotopic fractionation or isotope-selective
photodissociation, yield a n(C)-ratio of $58\pm 6$.  \citeN{KSK98} conclude
that, in contrast to the model results by \citeN{LPR93} the importance of
chemical isotopic fractionation is (almost) compensated for by the
isotope-selective photodissociation. Whether this compensation is peculiar
to the special conditions of the Orion Bar, or whether it also holds for
other massive star-forming cores is a significant question when addressing
galactic chemical evolution, since our current understanding is almost
entirely based on the observed carbon monoxide isotopomer abundance ratios.
It is therefore important to extend the study of \THCI\ to other regions.

In this Letter, we report measurements of the \tpt\ transition of \CI\ at
809341.97 MHz and the F$=5/2-3/2$, F$=3/2-1/2$, and F$=3/2-3/2$ components
of the equivalent transition of \THCI\ at 809493.7 MHz, 809125.5 MHz, and
809121.3 MHz, respectively, towards three galactic star-froming regions
associated with strong IRAS sources and CO emission: G 333.0-0.4, part of
the large \HII\ complex RCW 106 and associated with a molecular cloud with
bright CO lines (\citeNP{GHS77}, \citeNP{BBV84}) at a distance of 4.2 kpc
\cite{SG70}; NGC 6334 A, a bright FIR continuum region \cite{MFS79}
associated with a bright PDR (\citeNP{BAM00}, \citeNP{KJL00}) located within
a giant molecular cloud at a distance of 1.7 kpc \cite{Nec78}; and G
351.6-1.3, a luminous (\Lbol$ \approx 10^6$\Lsol) compact \HII\ region
excited by an embedded O/B cluster \cite{MLF85} at a distance of 5 kpc
\cite{RGM72}. These regions were selected because of strong emission in the
\tpo\ transition of \CI\ at 492 GHz \cite{HBB99} from a list of 30
candidates of galactic \HII\ regions having strong CO (\citeNP{BBV84},
\citeNP{HL98}) and CS (\citeNP{BNM96}, \citeNP{LEW98}) emission lines.

\section{Observations}

The observation were made in the Austral winter of 2000 with the 800 GHz
receiver at at the Antarctic Submillimeter Telescope and Remote Observatory
(AST/RO, cf. \citeNP{SBB01}). At these frequencies the FWHM beam size is
$\approx 80$\arcsec. As backend we used an AOS with a velocity resolution of
0.25 \kms\ at the observed frequencies. The zenith opacities at 809 GHz at
the times of observation were $\approx 1.0\pm 0.2$ System temperatures on
average were between 11,000 K for G 333.0-0.4, 18,000 K for NGC 6334 A, and
25,000 K for G 351.6-1.4.  The \CI\ and \THCI\ lines were observed in the
lower sideband simultaneously with the CO 7--6 line in the upper sideband.
The spectra were calibrated \cite{SBB01} to the effective radiation
temperature \TRST\ scale.  The frequencies reported by \citeN{KLS98} (cf.
Cologne Database for Molecular Spectroscopy -- www.cdms.de) were used to
determine the line centroids.

\begin{figure}
  \epsfig{figure=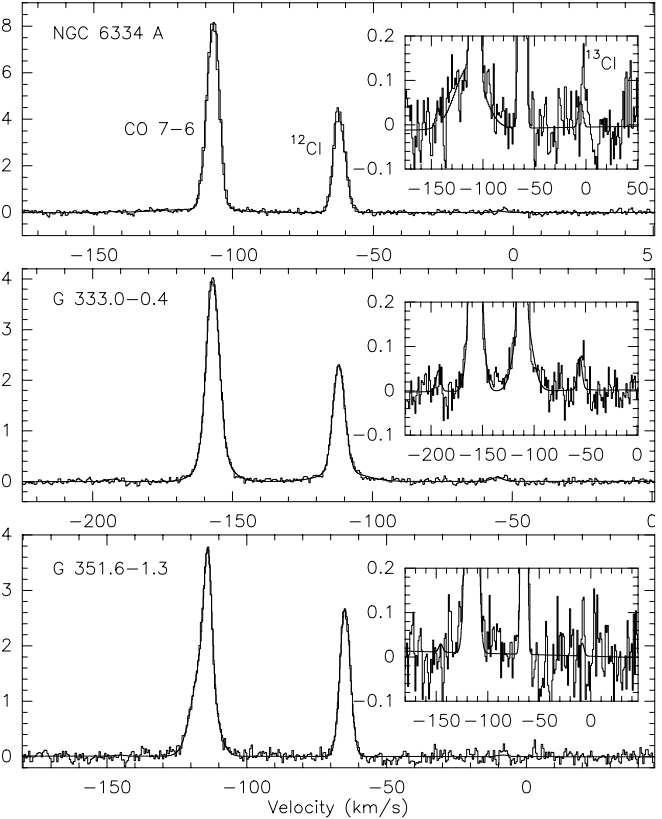,width=\columnwidth}
    \caption{Broad band spectra (50 hrs of integration time on source)
      towards the galactic star-forming
      regions NGC 6634 A (top), G 333.0-0.4 (middle) and G 351.6-1.3
      (bottom).  The insets show the same spectra magnified in y-scale. The
      \CI\ and \THCI\ lines were observed in the lower sideband
      simultaneously with the CO 7--6 line in the upper sideband.
      Note the erratic baseline toward the \THCI\ line emission in G
      351.6-1.3, which resulted in a skewed baseline fit and precludes a
      clear 3$\sigma$ assignment of this peak.} 
\end{figure}

\section{Results}

Fig.~1 shows the spectra obtained towards the three regions.  The strongest
HF component of \THC\ at 809 GHz, F$=5/2-3/2$, is clearly visible in the G
333.0-0.4 and NGC 6334 A spectra; it is marginally detected in G 351.6-1.3.
The weaker HF structure satellites, F=$3/2-1/2$ and F=$3/2-3/2$, are still
hidden in the noise in all cases.  Following the first detection towards
Orion by \citeN{KSK98}, these regions are the only other \THC\ detections to
date.  We derive the \CI/\THCI\ intensity ratio by a simultaneous Gaussian
fit of the lines (including {\it all} \THCI\ HF components) with {\it fixed
  spacing and a single line width}, and with the ratio of the \THCI\ 
amplitudes fixed to their quantum mechanical values of 0.600:0.333:0.067;
the free fit parameters are the common width and LSR-velocity, a common
amplitude and the relative line intensity ratio
$\alpha=\frac{I_{12}}{I_{13}}$. We included the CO 7--6 line from the other
sideband as an additional, independent Gaussian.  Due to broad line wings
apparent in the CO and \CI\ emission lines towards G 333.0-0.4, we fit two
line components (broad \& narrow) for all emission lines here. In the case
of NGC 6334 A and G 351.6-1.4, where no wings are apparent for the \CI\ line
but asymmetric line profiles are conspicuous for the CO 7--6 line, we fit a
second broad line component to the CO emission line only.  The fit results
for all sources are given in Table 1. The uncertainties quoted are the
formal fit errors based on a 1$\sigma$ excursion. Since the \THCI\ and \CI\ 
lines were measured simultaneously in the same receiver sideband,
calibration uncertainties are negligible.

\begin{table}
  \label{fits}
   \caption{Line fit results}
   \begin{flushleft}
   \begin{tabular}{|ccccc|}
  \hline
emission  & ${T_R}^*$ & $\Delta$v$_{\rm lsr}$ & $\alpha$ & \\
line      & [K]       & [kms$^{-1}$] & $\frac{I_{12}}{I_{13}}$ & \\
  \hline
  \multicolumn{5}{|c|}{NGC 6334 A} \\
  \hline
$^{13}$C{\sc i} & $0.10\pm 0.02$   & $4.9\pm 0.2$ & $45\pm 11$ & n\\
CO              & $8.04\pm 0.06$   & $5.3\pm 0.1$ &            & n\\
CO              & $0.14\pm 0.03$   & $32\pm 7$    &            & b\\
  \hline
  \multicolumn{5}{|c|}{G 333.0-0.4} \\
  \hline
$^{13}$C{\sc i} & $0.12\pm 0.02$   & $5.0\pm 0.1$  & $18\pm 1$ & narrow\\
$^{13}$C{\sc i} & $0.04\pm 0.01$   & $18\pm 1$     & $45\pm 4$ & broad\\
CO              & $3.57\pm 0.02$   & $5.4\pm 0.1$  &           & n\\
CO              & $0.46\pm 0.02$   & $12.6\pm 0.4$ &           & b\\
  \hline
  \multicolumn{5}{|c|}{G 351.6-1.3} \\
  \hline
$^{13}$C{\sc i} & $0.05\pm 0.01$  & $4.4\pm 0.1$  & $55\pm 10$  & n\\
CO              & $2.15\pm 0.02$  & $3.3\pm 0.1$  &            & n\\
CO              & $1.77\pm 0.02$  & $8.9\pm 0.2$  &            & b\\
    \hline
    \end{tabular}
    The $^{13}$C{\sc i} amplitudes given are the sum over the three HF
    components (see text).
    \end{flushleft}
\end{table}
In the optically thin case, the N(C)-ratio is directly given by the line
intensity ratio \al; beyond this limit an optical depth correction must be
applied.  We prefer this approach to the alternative of including the
optical depth into the fit by fitting saturated line profiles of the form
$1-{\mathrm{exp}}(-\tau \times \phi_\nu)$, because of the macro-turbulent
nature of the velocity field in the ISM and the corresponding failure to
observe saturated profiles even for optically thick lines of \THCO\ in any
astronomical source. This is supported by the Gaussian profile fit to the CO
7--6 line with a width only marginally wider than that of \CI.

\subsection{Optical Depth Estimate}

To derive the intrinsic \CI/\THCI\ column density ratio, N(C), we apply an
optical depth correction to \al.  Following \citeN{KSK98}, the ratio is
given by
\begin{\eq}
  \frac{N(^{12}C)}{N(^{13}C)} = \frac{\alpha}{\beta(\tau_{^{12}C})},
\end{\eq}
with the line escape probability $ \beta(\tau) = (1-e^{-\tau})/\tau$. To
further constrain the properties of the C{\sc i} emitting gas, we use the
line intensity ratio of the \CI\ \tpt\ and \tpo\ emission lines.  The 492
GHz \CI\ \tpo\ line transition has also been measured \cite{HBB99} towards
these sources; assuming that the emission is extended and need not be
corrected for different beam sizes at 492 and 809 GHz, we derive a line
intensity ratio of 0.48 for G 333.0-0.4, 0.63 for NGC 6334, and 0.75 for G
351.6-1.4. If the sources are centrally peaked, coupling to the narrower 809
GHz beam would be higher, and the intrinsic line ratios would be
correspondingly lower.

We use a single component escape probability excitation code \cite{SW85} to
interpret the observed line intensities.  As an additional constraint we
need an estimate of the source density. Only for NGC 6334 do we have an
estimate: \citeN{KJL98} compared their observed line intensities of \CII\ 
(158\micron) and \OI\ (146\micron \, \& 63\micron) and the CO line and FIR
continuum intensities to PDR models of \citeN{WTH90} to estimate the gas
density and FUV field towards the embedded radio continuum sources. Towards
NGC 6334 A they derive an average FUV field of ${\mathrm{log}\,} G_0 = 4.5$
and a total gas density of n(\MOLH) $= 6.3\times 10^3$ \perccm\ within their
1\arcmin\ beam. It is fair to assume that the density of the C{\sc i}
emitting gas in NGC 6334 A and the other two sources is similar or higher,
as it is located deeper in the molecular cloud. Densities of that magnitude
are also supported by the bright CO 7--6 emissions we observe.

As is shown in Fig.~2, this density immediately constrains the temperature
of the C{\sc i} emitting region to about 40 K or lower: at higher
temperatures, the \CI\ \tpt/\tpo\ line ratio at these densities would be
substantially higher than observed.  This temperature estimate is perfectly
consistent with the scenario of extended PDR emission.  In this regime of
density and temperature, the observed \CI\ \tpt/\tpo\ line ratio constrains
the column density per unit velocity interval to values below $2 \times
10^{17}$ \perscm/\kms\ and corresponding optical depths of the \CI\ \tpt\ 
line below $\tau \approx 0.4$. At higher column densities trapping would
push the line ratio to much higher values than observed.
\begin{figure}
  \epsfig{figure=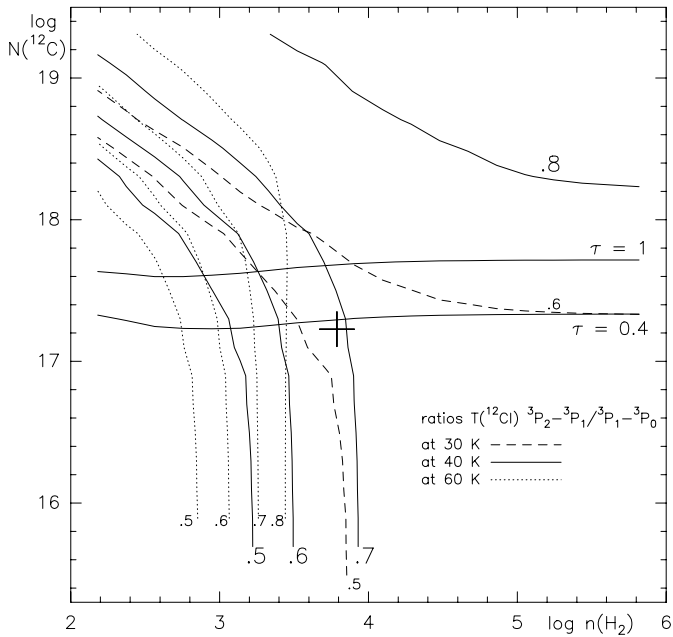,angle=-90,width=\columnwidth}
 \caption{Results from the line escape probability code: Bold lines show 
   $^{12}$C{\sc i} line ratios for 40 K.  At these temperatures and lower
   (30 K, dashed), densities of n(\MOLH) $= 6.3\times 10^3$ cm$^{-3}$
   indicate an optical depth of $\tau < 0.4$ and constrain the densities to
   N($^{12}$C) $< 2 \times 10^{17} \, {\mathrm{cm}}^{-2}/ {\mathrm{km \,
       s}}^{-1}$ (indicated by a cross).  At higher temperatures (60 K,
   dotted) the ratios at these densities would be substantially higher than
   observed.}
\end{figure}

The total column density in the $\approx 5$\kms\ wide \CI\ lines is then
about $10^{18}$\perscm; this corresponds, with the standard gas phase
abundance of carbon of about $10^{-4}$, to a hydrogen column density of
$10^{22}$\perscm.  Considering that each PDR surface has a C{\sc i} column
density corresponding to an $A_V$ of a few, we therefore conclude that 
we see a few PDR surfaces per beam, consistent with the usual scenario of 
a clumpy, UV-penetrated massive cloud core.

\section{Discussion}

For the range of density, temperature and column density derived above, our
radiative transfer model predicts line brightnesses for the \CI\ \tpt\ lines 
of around $5 - 10$ K. Since the measured line brightnesses are a few degrees in
the observed sources, we estimate a beam filling factor close to unity, 
consistent with extended, smooth emission within the AST/RO beam.  This scenario
suggests the \CI\ emission is moderately optically thin.  For the rough
upper limit of the optical depth $\tau \approx 0.4$, the optical depth
correction factor $1/\beta(\tau) = 1.21 $ only raises the N(C)-ratio derived
above by at most 25\%.  We estimate the N(C)-ratio in G 333.0-0.4 to be
$23\pm 1$ or lower ($56\pm 5$ for the weak extended component), that in NGC
6334 A to be $56\pm 14$ or lower, and that in G 351.6-0.4 to be $69\pm 12$
or lower.

The N(C)-ratios towards G 351.6-1.4 and NGC 6334 A are consistent with the
intrinsic isotopic ratios and do not call for significant isotopic
fractionation or isotope-selective photodissociation. These results confirm
the conclusion of \citeN{KSK98} for their Orion Bar data.  Towards G
333.0-0.4, however, we find the ratio to be substantially lower than the
average values derived from CO isotopomers.  In this context one should
recall, that \cite{LP93} find an increase in the CO/\THCO-isotopomer ratio
towards clouds exposed to higher UV radiation fields, supporting the
suggestion of PDR models that the depletion in \THCO\ can be ascribed to
less effective self-shielding and enhanced isotope-selective
photodissociation \cite{DB88}. This would then yield an increase in the
abundance of the \THC\ isotope, as is shown in the PDR-models by
\citeN{KSS94}, if not counter-balanced by chemical isotopic fractionation.
As chemical isotopic fractionation becomes more efficient with decreasing
temperatures and eventually dominates over isotope-selective
photodissociation, there will be a decrease in the abundance of the \THC\ 
isotope. This is shown in the PDR models of \citeN{LPR93}, which produce a
lower temperature in the transition zone due to reduced heating.  Cloud
surface temperatures therefore influence the balance of the two competing
effects and, ultimately, determine the carbon isotope abundance ratio.  The
temperature structure in PDR surfaces, details of which are not consistently
explained by present PDR models, might be responsible for the very different
atomic carbon isotope abuncance ratios observed to date.

\begin{acknowledgements}
  
  The Letter is dedicated to Rodney Marks, who was working
  as the Winterover Scientist for the AST/RO project when he died on May
  12$^{\mathrm{th}}$, 2000, during preparations for these observations.
  The CARA winter-over crew, Gene Davidson, Greg Griffin, David Pernic, and
  John Yamasaki, continued AST/RO operations in tribute to Rodney's memory,
  allowing these observations to be made.  The Universit\"at K\"oln
  contribution to AST/RO was supported by special funding from the Science
  Ministry of the Land Nordrhein-Westfalen and by the Deutsche
  Forschungsgemeinschaft through grant SFB 301.  This work was supported in
  part by United States National Science Foundation grant DPP88-18384, and
  by the Center for Astrophysical Research in Antarctica (CARA) and the NSF
  under Cooperative Agreement OPP89-20223.
  \end{acknowledgements}

\end{document}